 \documentclass[manuscript]{aastex}

\shorttitle{The Collimated Wind in NGC~253}
\shortauthors{Sugai, Davies, Ward}

\begin{document}


\title{The Collimated Wind in NGC~253}


\author{H. Sugai}
\affil{Department of Astronomy,
Kyoto University, Sakyo-ku, Kyoto 606-8502, Japan}
\email{sugai@kusastro.kyoto-u.ac.jp}

\author{R. I. Davies}
\affil{Max-Planck-Institut f\"ur extraterrestrische Physik,
Postfach 1312, Garching, Germany}
\email{davies@mpe.mpg.de}

\and

\author{M. J. Ward}
\affil{Department of Physics and Astronomy,
University of Leicester, University Road, Leicester LE1 7RH, UK}
\email{mjw@star.le.ac.uk}



\begin{abstract}
Near-infrared Fabry-Perot imaging has revealed
H$_2$ emission extended to
about 130 pc from the disk of NGC 253.
It is closely related to the hot plasma
observed in soft X-rays: filamentary
H$_2$ features are found at the {\it edges} of the hot plasma.
These are the places of {\it direct} interaction between
a superwind and its surrounding molecular gas.
We suggest that the filamentary
features actually trace a more or less conical shell-like
structure,
whose tangential line of sight to us is intensely observed.
The H$_2$ emission shell is most likely from the
molecular gas blown out
or swept to the side
by the hot plasma outflow.
Dust is associated with this molecular gas structure.
The outflow is tilted with respect to the disk,
possibly suggesting the inhomogeneous nature of the
interstellar medium in which the starburst takes place.
\end{abstract}




\section{Introduction}

NGC 253 has an outflow caused by starbursts in
its central regions \citep{str00,pie00,wea02,for00}.
Being almost edge-on and one of the closest starburst galaxies,
with a distance of $\sim 2.6$ Mpc \citep{puc88},
it is one of the best examples for studying the
structure of such outflows.
\citet{gar00} found SiO emission components
whose velocities imply that the molecular gas is 
flowing out of the disk.
The SiO emission is most likely caused by shocks.
The H$_2$ emission also traces the excited molecular gas
(e.g, Sugai et al. 2000, 1999, 1997), including shocked gas.
Tanaka et al.'s (1994) results showed that a search
for extended H$_2$ emission would be fruitful,
although their spatial resolution  
was not enough to discuss its morphology. 
In this Letter
we report our high spatial resolution Fabry-Perot 
imaging from UKIRT of the H$_2 \ v=1-0$ S(1) emission in this galaxy.
We also show a {\it Hubble Space Telescope} narrow-band 
image, which was
obtained from archive data (Prop ID 7218: Rieke, M.),
to emphasize 
the apparently sharply delineated components of the extended 
H$_2$ emission.

\section{Observations and Data Analysis}

We obtained line images in the H$_2 \ v=1-0$ S(1)
transition
using a Fabry-Perot imager mounted on the Cassegrain focus of
the UKIRT 3.8 m telescope
on Mauna Kea Hawaii on 1996 December 19.
A {\it K}-band etalon with 50 mm effective aperture made by Queensgate
Instruments was set in a collimated beam.
A cooled narrow band filter (0.9\%)
was used for sorting one order of interference.
We used the near-infrared camera, IRCAM3, which has a 256 $\times$ 256
InSb array, with a pixel scale of $0^{\prime\prime}.3$.
Wavelength calibration was carried out using a Krypton lamp.
The spectral resolution was 325 km s$^{-1}$.
The peak transmission at the target object position was set to 
its systemic velocity, $v_{helio} \sim 245$ km s$^{-1}$
(e.g., da Costa et al. 1991).
Although the Fabry-Perot provided an unvignetted field of 
approximately $1^{\prime}$ in diameter for this velocity,
the rotation of NGC 253 may have caused loss of some H$_2$ 
flux in its receding southwestern edge.
However, this only affects the farthest part of the image
and does not alter the discussion in this Letter.
For each line image,
we repeated the following observing sequence:
(1) line frame, (2) continuum frame, (3) line frame again,
and (4) continuum frame on the opposite side of the line.  
The same observing sequence was carried out for sky frames.
Two sets of such sequences provided a total on-line integration 
time of 12 minutes on the source. 
After sky subtraction from each object frame by the corresponding
wavelength sky frame,
we carried out flat fielding by using sky frames at the same
wavelength setting.
A transmittance correction, which included atmospheric absorption
and the transmittance of the order-sorting filters,
was made using the A3 V standard star BS~232 observed at a similar
time and air mass and at the same wavelength setting.  

We also retrieved {\it HST} NICMOS archive data.
Images through narrow-band filters F212N 
($\lambda_c=2.1213 \mu$m; $\lambda / \Delta\lambda=1 \%$)
and F215N 
($\lambda_c=2.1487 \mu$m; $\lambda / \Delta\lambda=1 \%$)
were used to create a H$_2 \ v=1-0$ S(1) line image, 
although the continuum subtraction
is less accurate than the Fabry-Perot data.
We also use {\it HST} WFPC2 archive data of F814W broad-band filter
imaging.




\section{Morphology of Extended H$_2$ Emission}

Figure~\ref{h2cont} shows 
the spatial distribution 
of the H$_2$ emission obtained with the UKIRT Fabry-Perot.
Strong H$_2$ emission is associated with the continuum
emission, which traces 
the disk stellar distribution.
However, more extended H$_2$ emission is also detected. 
Features found in our Fabry-Perot data are 
consistently seen also in the {\it HST} NICMOS image.
We call the apparently ``filamentary'' features A, B, 
C, and D.
Features B and D are more clearly recognized in the 
{\it HST} image 
since the contrast of these sharp features against nearby H$_2$
emission is higher.
Feature A extends southeast 
to a height of at least $10^{\prime\prime}$.
The apparently narrow features A and B, and possibly also C,
are tightly related to the intense hot plasma 
observed by \citet{str00} and \citet{wea02} in soft (0.2--1.5 keV) 
X-rays: these H$_2$ emission features are 
found at the {\it edges} of the hot plasma 
(Figure~\ref{x}).
They trace the
expected places for an interaction between a superwind
and its surrounding molecular gas.
\citet{str00} showed that the X-ray emitting gas distribution 
is consistent with a conical outflow, although it may depend
on the observed energy band whether or not the conical outflow
has a hollow structure \citep{pie01}. 
Since it is unlikely that the filaments accidentally occur
exactly and only at our tangential line of
sight to the hot plasma outflow,
we propose that (at our spatial resolution) the H$_2$ emission
structures consist of a more or less bi-conical 
shell-like structure around the outflow; and that the
^^ ^^ filaments" in fact trace the tangential line of sight to
this H$_2$ emission shell.
Although the shell may be inhomogeneous and, for example, may 
actually consist of many unresolved filaments embedded in a less dense
sheet, even in this case it is unlikely
that those filaments exist only at our line of sight.

The hot plasma outflow is tilted with respect to
the axis perpendicular to the disk, 
when it leaves the disk plane, 
as found in the outflow of M82 by \citet{sho98}.
The position angles of ^^ ^^ filamentary" features A, B, C, and D
on the plane of the sky are offset from the disk minor axis direction
roughly by $\sim 12^{\circ}$, $\sim 37^{\circ}$, $\sim 25^{\circ}$,
and $\sim 3^{\circ}$, respectively, towards the northeast side 
of the disk.
The apparent offset seems larger on the northeast side, i.e.,
for features B and C, compared with features A and D.
The tilt of the outflow may relate to
the inhomogeneous nature of the interstellar medium in
which the starburst takes place. 
\citet{str02} discusses why the interstellar medium must be 
inhomogeneous in the vicinity of the starburst. 
If it was uniform dense gas,
most of the initial kinetic energy of individual supernova
remnants and stellar-wind-blown bubbles would be lost to
radiation before being used for the heating to produce
a superbubble.

\section{Origin of Extended H$_2$ Gas} \label{ori}

From where does the H$_2$ molecular gas in features A, 
B, C, and D originate?
There are four possibilities: 
(1) H$_2$ reproduction from the hot plasma, 
(2) molecular gas blown out by superwinds, 
(3) molecular gas at the same height but that has been swept to the
side as the superwind expanded, and
(4) molecular gas which was already where it is now seen.
In Figure~\ref{x}, 
the H$_2$ emission is found only 
around the beginning of the hot plasma outflow and is 
not found further out.
This rules out possibility (1) because
H$_2$ reproduction would be seen almost everywhere along the hot
plasma edges.
Figure~\ref{hst814} shows an optical image from 
{\it HST} WFPC2 archive data.
Registration of this image to the near-infrared images
was achieved using coordinates determined by  \citet{wat96}.
We use this F814W image to investigate the dust distribution,
because optical images are more sensitive to dust extinction
than near-infrared images.
We find a bright region coincident with the X-ray plasma and
apparently ^^ ^^ narrow" lanes at the positions 
of H$_2$ emission features A and (less clearly) B. 
This suggests that dust is lacking in the hot plasma, and instead is
tightly associated at least with features A and B, although
it might extend farther from the disk plane
than the H$_2$ emission features.
Less dust exists just outside the ^^ ^^ narrow" dust 
lanes, suggesting that 
it may form a 
conical shell-like structure similar to the H$_2$ emitting gas.
This makes possibility (4) less likely.  
If dust had existed there previously, its distribution outside the hot
plasma would be more homogeneous.
The ^^ ^^ narrow" lanes imply that the dust has been 
blown out or swept to the side 
in a similar way to the H$_2$ emitting gas.
The stronger H$_2$ emission feature A appears to be located
on the northeastern side (i.e. inside) of the corresponding dust
lane, suggesting a stronger 
interaction with the superwind inside the dust shell. 

Although from our observations alone we cannot conclude which of 
(2) and (3) is the origin of the extended H$_2$ gas,
results of theoretical works provide some insight for this
(e.g., Strickland \& Stevens 2000; Williams \& Dyson 2002).
For example,
\citet{suc94} 
suggested that the wind drags out material from the disk.
The extended H$_2$ gas in NGC~253 might correspond to this 
dragged-out material, which may have a contribution for
collimating the wind.
It is also possible that the H$_2$ gas is being entrained
by the wind while dense molecular gas remaining in the disk 
collimates the wind.
Even in this case, the base of the extended H$_2$ gas closely
traces the place of wind collimation.

Observations of SiO emission by \citet{gar00}
suggest that two SiO filaments come out of
the plane of the nucleus of NGC 253.
According to these authors, the velocities at the filament positions
are easily accounted for if the molecular gas is outflowing.
This implies that it is actually possible that a significant
amount of molecular gas 
is coming out from the disk plane, although the positional 
difference between the SiO filaments and the extended H$_2$ 
emission features implies that  
they are not the same phenomenon.
Kinetic energies required for starburst
outflows jump up by 2-3 orders of magnitude after the detection of 
dust and molecules, compared with those required only for ionized gas.
This is possibly because there is so much mass associated with
molecular gas phase compared to any other detectable phase.
A recent example of this is discussed by 
\citet{bla03}
in the Galactic Center.

\section{What Excites the Extended H$_2$ Gas?}

\citet{tan94} found spatial variation of the
H$_2 \ v=2-1$ S(1) / H$_2 \ v=1-0$ S(1) line ratio
over a half arcminute from the nucleus, although they could not
resolve the extended H$_2$ emission reported in this Letter.
But their results may imply spatial variation of the H$_2$
excitation mechanisms.
From slit spectroscopy, \citet{eng98} suggested 
that 25 \% of the H$_2 \ v=1-0$ S(1) emission 
in a central $2^{\prime\prime}.4 \times 12^{\prime\prime}$
aperture is excited by fluorescence -- but this applies only to 
the strong nuclear H$_2$ emission.
Ideally, spectroscopy exactly at the extended H$_2$
emission features would be needed to find the excitation
mechanisms at the interface between the hot plasma
and the molecular gas.
However, a discussion of the energetics 
places some constraints on its excitation mechanisms.
We measure an H$_2$ luminosity of order 
$\sim 1 \times 10^{37}$ erg s$^{-1}$ for feature A.

First we consider X-ray heating.
While \citet{wea02} derived an unabsorbed 2--10 keV luminosity
of $\ge 10^{39}$ erg s$^{-1}$ for a heavily absorbed source of 
hard X-rays embedded within the nuclear starburst region,
\citet{pie00} obtained an X-ray luminosity of
$9 \times 10^{38}$ erg s$^{-1}$ for the X-ray plume.
Assuming the efficiency of converting X-rays into 
H$_2 \ v=1-0$ S(1) emission to be $\sim 0.5 \%$ through X-ray heating
(see Draine and Woods 1990),
it seems difficult for the X-ray plume alone to provide the energy input
for the H$_2$ emission because only a small fraction ($\sim$ 0.1) of 
its X-rays would be incident on the solid angle around feature A.
However, until an upper limit to the
unabsorbed nuclear luminosity is established, 
we cannot rule out the possibility that the nuclear source
itself might supply enough X-rays.

Next we consider UV fluorescence.
\citet{eng98} derived an extinction corrected
ionizing photon number of $\sim 1 \times 10^{53}$ s$^{-1}$
from the starbursts.
Taking into consideration an upper limit of 37000 K for the exciting stars
\citep{eng98}
and also Table 5 of Puxley et al. (1990), we find that
of order $10^{37}$ erg s$^{-1}$ of H$_2 \ v=1-0$ S(1) emission is 
obtained when integrated over the {\it whole} sphere of radius 
$\sim 10^2$~pc
(for photodissociation region gas density less than
$10^3$ cm$^{-3}$).
We used a model for clustered stars residing within a communal
HII region, in which the fluorescent H$_2$ emission originates from
the edge of the giant HII region. 
Models in which individual stars have their own fluorescent H$_2$
emission regions cannot produce large-scale structures such as
feature A.
Since only of order one tenth of the UV flux from the starbursts in
the central regions would 
reach the solid angle around feature A,
we conclude this is not enough
unless dense photodissociation region
($> 10^3$ cm$^{-3}$) has a covering factor of an order of unity.
Although this may be unlikely, we cannot rule out a contribution 
from the UV fluorescence particularly to
excitation of those parts of the shells close to the disk.

Shocks may more easily produce the observed H$_2$ emission.
The H$_2$ surface brightness at feature A is of order
$\sim 10^{-15}$ erg s$^{-1}$ cm$^{-2}$ arcsec$^{-2}$.
This is easily achieved by shock models
(e.g., Burton et al. 1990).
{\it If} we assume that, like the SiO emission, the H$_2$ components are
shock excited, 
we can estimate the mass of the hot molecular gas
to be $\sim 7$--5 {\it M}$_\odot$ for each of features A and B.
Although it is likely that a much larger mass of cold molecular
gas is associated with this hot gas, the
hot gas mass itself is at least much smaller than
outflowing gas/dust masses found so far:
the gas mass for the SiO southern filament is estimated as
$\sim 2.0 \times 10^6$ {\it M}$_\odot$
\citep{gar00} while the dust mass for the same filament
is estimated as
$\sim (0.4$-$2) \times 10^4$ {\it M}$_\odot$
\citep{gar00,alt99},
when the distance of 2.6 Mpc is assumed.
Therefore, it is possible that the hot molecular gas could have been
blown out from the nuclear regions of NGC 253.

\section{Similarity to IC 694}

The more distant and luminous starburst galaxy IC~694 also exhibits
extended H$_2$ emission similar to that in NGC~253.
It has a bipolar (or butterfly-like) structure in its
central regions \citep{sug99,alo00}.
The distance from the nucleus of IC  694 to the starting point
of this structure is $\sim 100$ pc 
($\sim 0^{\prime\prime}.5$),
comparable to NGC 253
($\sim 100$ pc and $\sim 70$ pc for features A and B, respectively).
The structure extends to about 300 pc from the disk.
These characteristics are also apparent in NGC 253, although the
case of IC 694 seems rather larger.
Although even {\it Chandra} may not be able to clearly 
resolve the nuclear X-ray emission from the hot plasma in
the central regions of IC~694, the H$_2$ emission
might, as for NGC~253, trace the edges of the hot plasma.
If so, then we can glean information
about the intense hot plasma distribution at the root of the 
superwind without seeing it directly.

\acknowledgments

We thank the staff at UKIRT and the JAC for their help during
the observing run.
The United Kingdom Infrared Telescope is operated by the 
Joint Astronomy Centre on behalf of the U.K. Particle Physics
and Astronomy Research Council.
We thank the referee for helpful comments on improvements to
the paper.

\clearpage


\begin{figure}
\caption{Spatial distribution of the H$_2 \ v=1-0$ S(1) emission. 
The continuum emission is subtracted.
Left: From UKIRT Fabry-Perot imaging.
The colours are re-scaled inside 
to show the
morphology of the stronger emission.
Labels are for continuum peaks, 
which were named
by \citet{kal94}.
Right: {\em HST} NICMOS archive data.
\label{h2cont}}
\end{figure}

\begin{figure}
\caption{
Left: From \citet{wea02}. Three-color composite {\it Chandra} X-ray image.
Red, yellow, and blue indicate the X-ray ^^ ^^ colors"
of 0.2-1.5 keV(soft), 1.5-4.5 keV (medium), and 4.5-8 keV (hard),
respectively.
The plus sign marks the position of the radio core
\citep{tur85}.
Right: With the {\it HST} NICMOS H$_2 \ v=1-0$ S(1) emission image
overlayed.
\label{x}}
\end{figure}

\begin{figure}
\caption{
Left: F814W image 
obtained from {\it HST} WFPC2 archive data.
Brighter regions have larger intensities.
Right: With the {\it HST} NICMOS H$_2 \ v=1-0$ S(1) emission image
overlayed.
\label{hst814}}
\end{figure}

\end{document}